\documentclass[%
 aip,
 amsmath,amssymb,
 reprint,%
]{revtex4-1}

\usepackage{graphicx}
\usepackage{dcolumn}
\usepackage{bm}

\usepackage[utf8]{inputenc}
\usepackage[T1]{fontenc}
\usepackage{mathptmx}
\usepackage{braket}

\begin{document}

\preprint{AIP/123-QED}

\title{Influence of nanostructuring on silicon vacancy center spins in diamond pillars}

\author{T. Lutz}
 \affiliation{ETH Z\"urich,Institute for Quantum Electronics, Otto Stern Weg 1, 8093 Z\"urich, Switzerland}
\author{T. Masuda}%
\affiliation{University of Calgary, Institute for Quantum Science and Technology and Department of Physics and Astronomy, 2500 University Drive NW, Calgary, Canada}

\author{J.P. Hadden}%
\affiliation{Cardiff University, School of Physics and Astronomy, 5 The Parade, Newport Road, Cardiff, United Kingdom}

\author{Ilja Fescenko}%
\affiliation{University of New Mexico, Center for High Technology Materials and Department of Physics and Astronomy, Albuquerque, USA}

\author{Victor Acosta}%
\affiliation{Center for High Technology Materials and Department of Physics and Astronomy, University of New Mexico, Albuquerque, USA}

\author{Wolfgang Tittel}%
\affiliation{University of Calgary, Institute for Quantum Science and Technology and Department of Physics and Astronomy, 2500 University Drive NW, Calgary, Canada}
\affiliation{Delft University of Technology, QuTech and Kavli Institute of Nanoscience, Lorentzweg 1, Delft, Netherlands}
\author{P.E. Barclay}%
\affiliation{University of Calgary, Institute for Quantum Science and Technology and Department of Physics and Astronomy, 2500 University Drive NW, Calgary, Canada}

\date{\today}

\begin{abstract}
Color centers in diamond micro and nano structures are under investigation for a plethora of applications. However,  obtaining high quality color centers in small structures is challenging, and little is known about how properties such as spin population lifetimes change during the transition from bulk to micro and nano structures. In this manuscript, we studied various ways to prepare diamond samples containing silicon vacancy centers and measured how population lifetimes of orbital states change in pillars as we varied their dimensions from approximately 1 $\mu$m to 120 nm. We also researched the influence of the properties of the diamond substrate and the implantation and annealing methods on the silicon vacancy inhomogeneous linewidth and orbital lifetime. Our measurements show that nominally identical diamond samples can display significantly distinct inhomogeneous broadening. We observed weak indications that restricted vibrational modes in small structures may extend population lifetimes. However, imperfections in the crystal lattice or surface damage caused by etching reduce population lifetimes, especially in the smallest structures.
\end{abstract}

\maketitle

Color centers in diamond are used for a broad variety of applications  such as sensing \cite{Maze2008}, hybrid optomechanical systems \cite{Burek2016, Golter2016, Lee2017} and quantum information processing \cite{Jelezko2004, Dutt2007, Fuchs2011, Hensen2015, Childress2013, Atature2018}. Many of these  applications require color centers implanted \cite{Evans2016,Ilja2017} within micro or nano structures \cite{Schroeder2016} that are engineered to modify the optical\cite{Fujita1296} or phonon density of states \cite{Lutz2016}, and they require color centers that are stable in their emission wavelength. The latter is particularly important in order to use the centers as sources of indistinguishable photons \cite{Sipahigil2014, Marseglia2018} in quantum networking applications. This also demands that an ensemble of color centers shows narrow inhomogeneous broadening.  However, no systematic studies on how the transition from bulk to nano-structures influences spectroscopic properties have been reported for color centers in nanostructured diamond. 

Among the numerous diamond color centers, the negatively charged  silicon-vacancy center (SiV) has recently received much attention \cite{Becher2019,Sipahigil2016, Sohn2017, Wein2017} due to its favorable spectroscopic properties \citep{becker2017} that make it appealing for the above mentioned applications. Its inversion-symmetric potential makes it less susceptible to spectral diffusion compared to the nitrogen-vacancy (NV) center, which exhibits a permanent electric dipole moment. Furthermore, more than 80\% of its emission is contained in its zero phonon line \citep{Neu2011}. It was for these reasons that the SiV was selected for the presented studies, however, we believe that the conclusions drawn here may also be relevant for other color centers.

For the experiments presented in this manuscript, we examined SiV color centers created by ion implantation in several diamond samples from the same supplier. We studied differences between the samples and investigated to what degree thermal annealing can alter their properties in terms of inhomogeneous linewidth and lifetime of orbital states. Furthermore, we investigated how the SiV orbital state lifetimes change when their environment is etched to create nano-pillars of varying size. We found that variations between diamond samples, possibly related to strain, can play an important role in obtaining SiV with a narrow inhomogeneous broadening. Re-annealing was not found to be successful in reducing the inhomogeneous broadening of implanted SiVs. Finally, we observed that orbital state lifetimes tend to decrease in the smallest nanostructures rather than increase, due to a suppression of resonant vibrational modes.\\\\


All diamond samples studied here were purchased from Element Six and were identical according to their specifications; 3$\times$3$\times$0.5 mm electronic grade (typical nitrogen concentration of $10^{13}$/cm$^3$ and NV concentration of $10^{10}-10^{11}$/cm$^3$) diamond chips, polished on both sides. The low NV concentration is required since the presence of many NVs in the sample can mask the SiV signal. The samples were cleaned using a mixture of perchloric, nitric, and sulphuric acid (1:1:1) heated to $\sim$ 250 $^{\circ}{\rm C}$ (tri-acid solution), followed by a 3:1 mixture of sulphuric acid and hydrogen peroxide (Piranha solution) pre-heated to 90 $^{\circ}{\rm C}$. After those cleaning steps, silicon ions ($^{28}$Si$^+$) were implanted into the sample with a dose of $10^{11}$ ions/cm$^2$ and an energy of 150 keV by Innovion Corporation. This implantation energy was predicted to result in an implantation depth of around 120 nm according to SRIM simulations\cite{Ziegler2010}. Subsequently, the samples were cleaned  again using the tri-acid solution and then annealed for 12 hours at temperatures up to 1100 $^{\circ}{\rm C}$ and under vacuum ($< 10^{-6}$ torr) \cite{Evans2016}, followed by another tri-acid cleaning step.

\begin{figure}[]
\centering
\includegraphics[width= 0.5 \columnwidth]{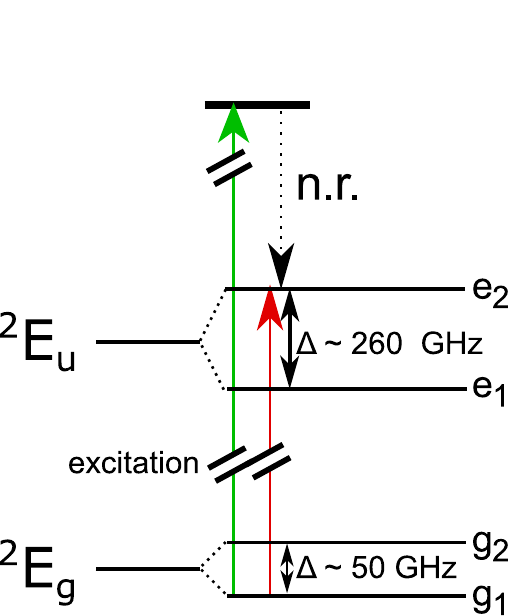}
\caption{Relevant energy-level structure of the negatively charged SiV. Off-resonant excitation (green arrow) populates the excited levels $e_{1}$ and $e_2$ via non radiative (n.r.) decays. Resonant excitation is represented by a red arrow.}
\label{fig:siv_levels}
\end{figure}

To investigate  the effect of nanofabrication on SiV properties, we first characterized SiVs implanted in the bulk samples after implantation and annealing. The relevant electronic level structure of the SiV is shown in Fig.~\ref{fig:siv_levels}. It is composed of the ground ($^2\rm E _g$) and excited state ($^2 \rm E _u$). For both, the spin-orbit interaction partially lifts the degeneracy and splits the states into the levels $g_1$, $g_2$ and $e_1$, $e_2$, respectively. Each of those levels further splits under the application of a magnetic field. The SiV features an optical transition at 737 nm (red arrow) that we used for our experiments. The diamond sample was cooled to 5 K using a Montana Instruments closed cycle cryostat (Nanoscale Workstation). A standard homebuilt confocal microscope, described in \cite{Hadden2017}, was employed for our spectroscopic measurements and SiV characterization: A long-working-distance microscope objective (NA = 0.5) outside the cyrostat and positioned using a three axis piezo stage was used to focus the excitation sources through the cyrostat top window. A single mode fiber was used as a pinhole to spatially filter the collected photoluminescence. This confocal microscope focused the green laser to a spot size of about 1 $\mu$m and always excited several color centers at the same time for the implantation density used here. Thus, all our bulk spectra reflect inhomogeneous broadening of the illuminated SiV. 

As a first step in SiV characterization, we excited the center incoherently with a 532 nm green laser (green arrow in Fig.~\ref{fig:siv_levels}) and monitored the emission using a spectrometer (Acton SP2750). 

\begin{figure}[]
\centering
\includegraphics[width=\columnwidth]{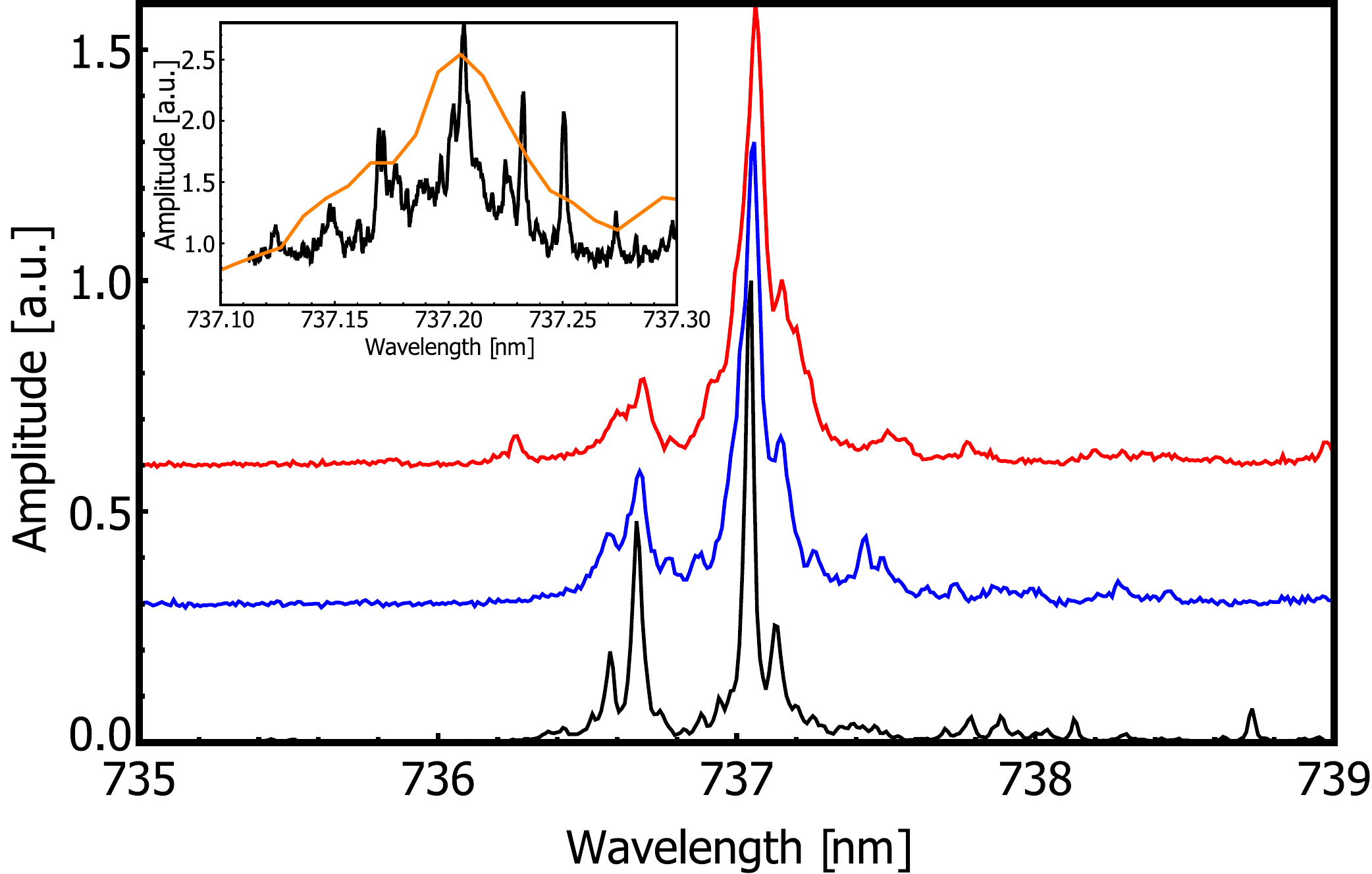}
\caption{Comparison of fluorescence emission spectra (532 nm excitation) of SiV ensembles created via implantation and annealing in different diamond chips under the same conditions. Black trace: sample 3, blue trace: sample 1, red trace: sample 2. Inset: Comparison of a resonant excitation photoluminescence  scan and lower-resolution fluorescence spectrum under off-resonant excitation for sample 3.}
\label{fig:singlecompare}
\end{figure}
Spectra from three different diamond chips, all produced with the same method, are shown in Fig.~\ref{fig:singlecompare}. Each spectrum shows peaks corresponding to the four expected optical transition lines connecting the two excited states with the two ground states. However, only sample 3 (represented by the black trace) features inhomogeneous broadening narrow enough so that each individual line is clearly resolvable. The resolution of the spectrometer ($\approx$ 20 GHz) was not sufficient to resolve the contributions of the individual SiV to the inhomogeneous broadening. To obtain high-resolution spectra, we scanned a tunable diode laser (New Focus Velocity TLB-6700) with a linewidth of around 1 MHz across the 737 nm absorption line, and monitored the  photoluminescence from the phonon sideband \cite{jahnke2015}. An excerpt of the high-resolution spectrum together with the corresponding part of the low-resolution spectrum from the spectrometer is shown in the inset of Fig.~\ref{fig:singlecompare}. 

The spectrum from sample 3 clearly shows contributions from SiVs with different emission wavelengths, likely due to each center experiencing a slightly different crystal environment in terms of residual stress from crystal polishing or implantation (i.e. the line is inhomogeneously broadened). However, since all samples were processed following the same implantation and annealing process, we conclude that the observed differences  in inhomogeneous linewidth are likely caused by differences in the original diamond chips. 

\begin{figure}[]
\centering
\includegraphics[width=\columnwidth]{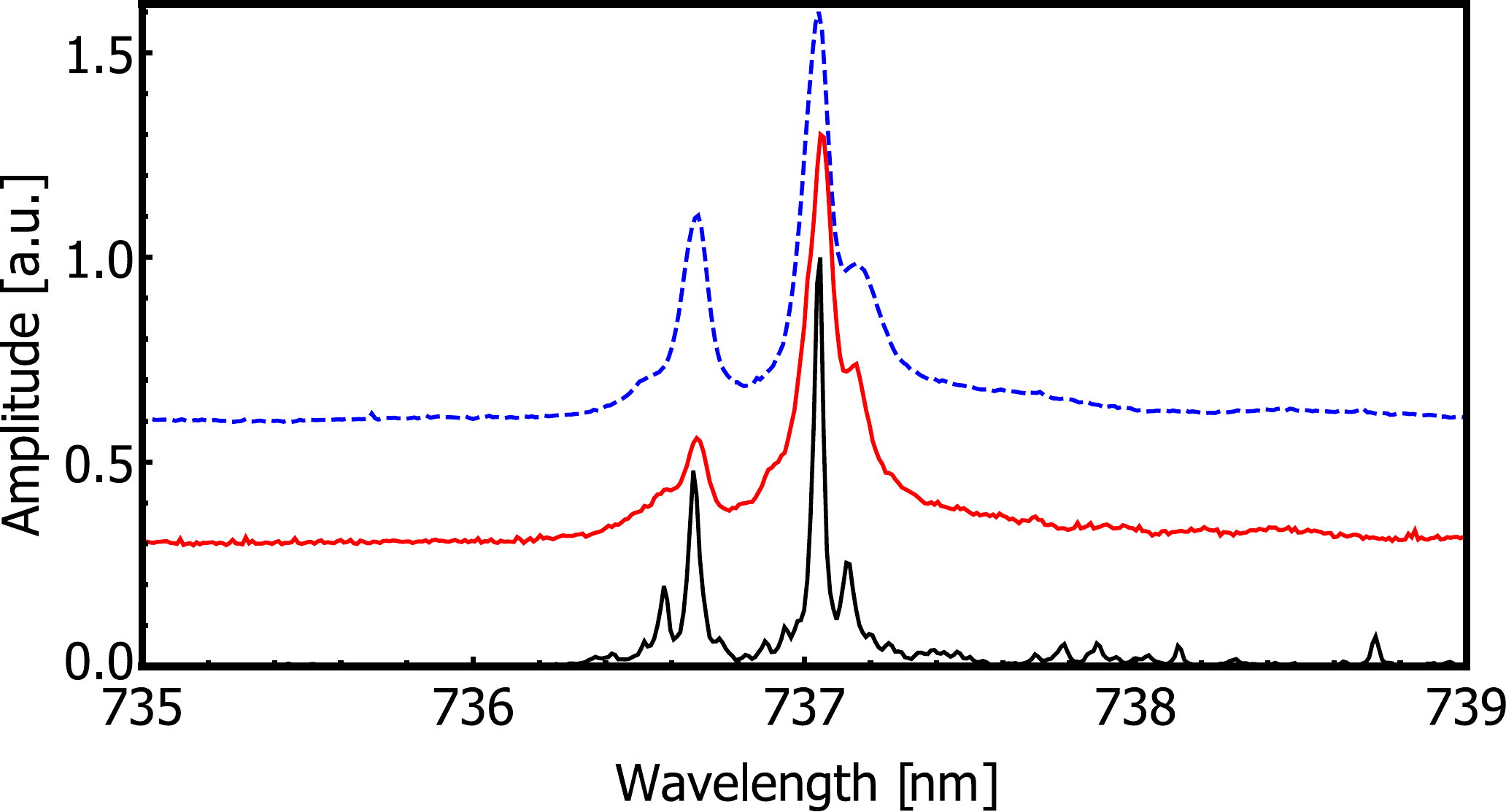}
\caption{High resolution resonant excitation photoluminescence  measurements of implanted SiVs, comparing the inhomogeneous broadening in sample 1 before (red trace) and after additional annealing (blue trace) for 12 hours at 1100 $^{\circ}{\rm C}$ and under vacuum ($<10^{-6}$ Torr). For reference, the spectrum showing the inhomogeneous broadening without additional annealing of sample 3 (black trace) is also shown.}
\label{fig:annealing}
\end{figure}

Since small inhomogeneous broadening is a key figure of merit, we investigated to what degree re-annealing the samples can help to relieve residual stress and reduce the inhomogeneous broadening of the centers. Sample 1 (see Fig.~\ref{fig:singlecompare}) was re-annealed in a home-built vacuum (<$10^{-6}$ torr) tube furnace. The furnace temperature was increased from 20-800 $^{\circ}{\rm C}$ over 6 hours, held at 800 $^{\circ}{\rm C}$ for 4 hours, ramped to 1100 $^{\circ}{\rm C}$ in 4 hours, and held at 1100 $^{\circ}{\rm C}$ for 12 hours. Figure~\ref{fig:annealing} shows the inhomogeneous broadening, averaged over measurements taken from many different spots on the sample, before annealing (red trace) and after annealing (red trace).
There is no observable difference between the spectra before and after annealing. Thus, thermal treatment beyond the initial annealing after ion implantation does not influence the amount of inhomogeneous broadening. 
Pre-etching the surface, i.e. removal of the top layer by reactive ion etching to remove any polishing damage, also did not reduce the inhomogeneous broadening. \\\\


For applications involving color centers embedded within nanostructures, it is valuable to understand how their properties are affected by the nanofabrication process as well as the resulting geometry. This is particularly interesting for the coupling of color centers to photonic or phononic nanocavities, where  predicted enhancements of the colour center's spin dynamics from locally engineered density of states can be degraded by deleterious effects resulting from etched surfaces and other nanofabrication-induced changes to the material. Ideally, such changes can be minimized through optimized processing, for example by using gentle etching processes \cite{Cui2014} and surface treatments \cite{fu2010,chu2014}. To investigate how nanofabrication, and in particular surfaces created using oxygen plasma etching, influences SiV properties, we conducted a study of the lifetime of the upper orbital state (labeled $g_2$ in Fig. \ref{fig:siv_levels}) of the negatively charged SiV ground state manifold. These measurements were conducted for SiV colour centres in nanostructures with varying dimension, as described below. We performed these experiments on sample 3, which showed the smallest inhomogeneous broadening in the characterization experiments described above.

\begin{figure}[h]
\centering
\includegraphics[width=\columnwidth]{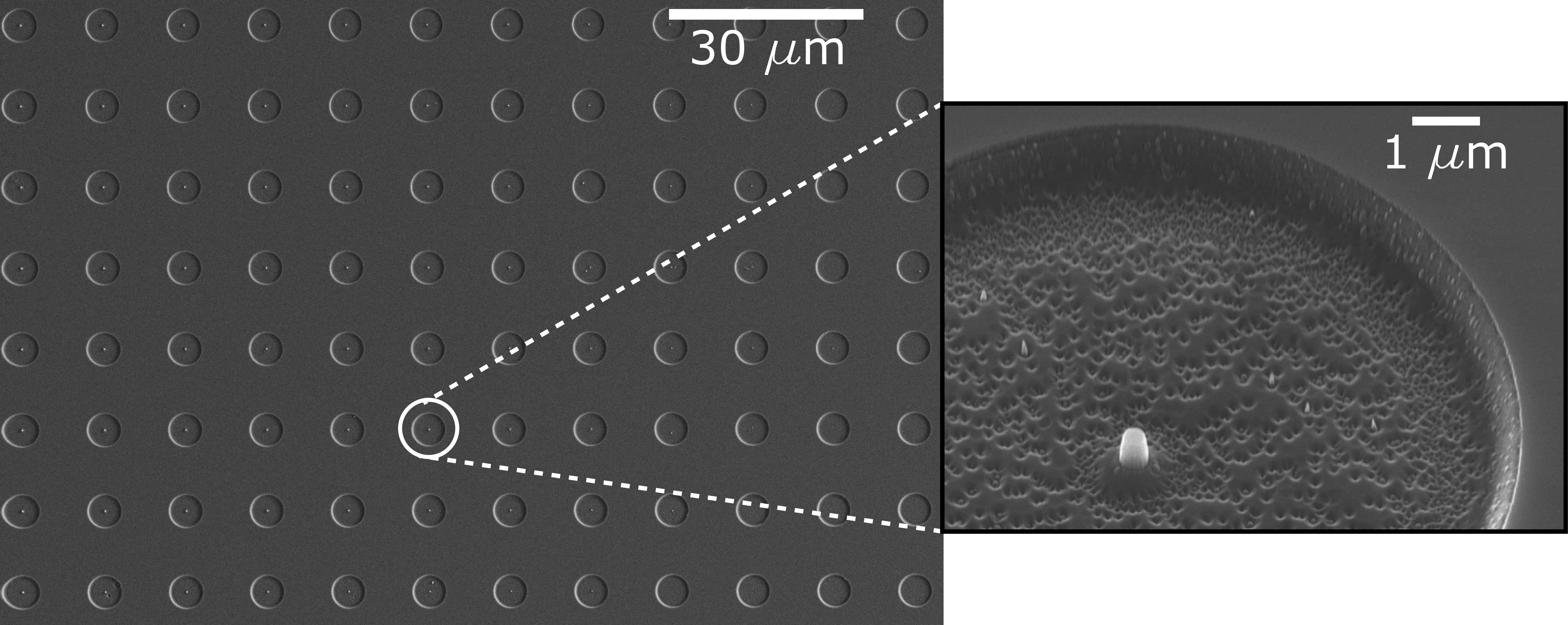}
\caption{Scanning electron microscope images of the fabricated nano-pillars. The inset image shows a 225 nm diameter pillar.}
\label{fig:sem}
\end{figure}

The nanostructure pattern used here to probe the influence of nanofabrication on SiV spin lifetime was an array of nanopillars with diameters varying from 120 to 900 nm  and heights of approximately 1 $\mu$m. The pillars were created using electron beam lithography followed by C$_4$F$_8$/SF$_6$ reactive ion etching to transfer the pillar array pattern to a silicon nitride hard mask deposited on the diamond chip using plasma enhanced chemical vapour deposition (PECVD). An oxygen plasma etch was then used to  transfer the pattern to the diamond chip. We followed the same process as described in \cite{Khanaliloo2015} omitting only the final isotropic undercutting step which was not required for these micropillar structures. After etching, the sample was cleaned in hydrofluoric acid solution to remove the SiN, followed by tri-acid and Piranha solutions. A scanning electron microscope image of the resulting nanopillars is shown in Fig.~\ref{fig:sem}.

\begin{figure}[]
\centering
\includegraphics[width=\columnwidth]{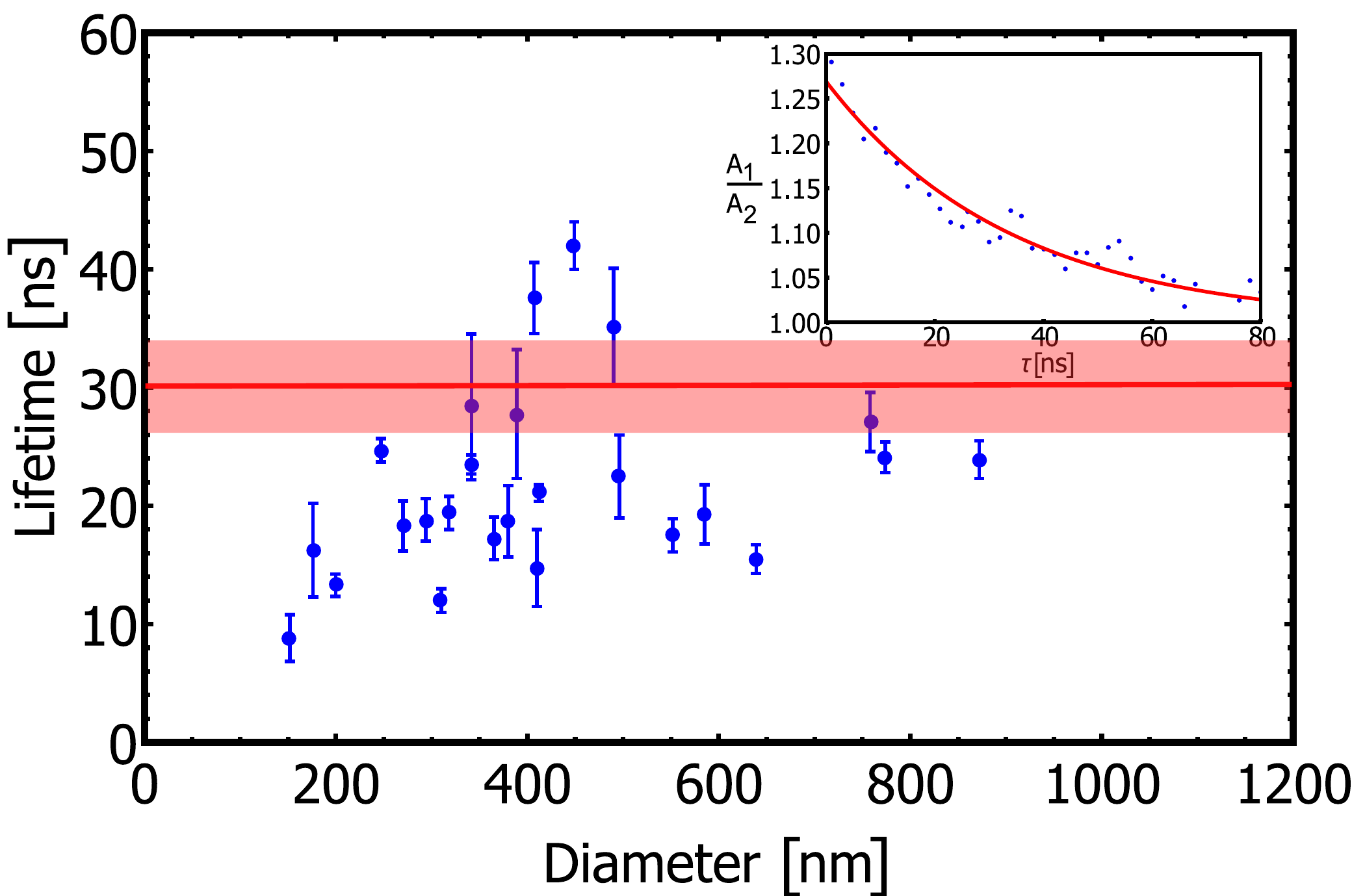}
\caption{Orbital state lifetime of SiV centers in nano-pillars of different diameters. Experimental data is displayed as blue dots. Each dot is an average over several nanopillars with the errorbar corresponding to the standard deviation of the measurements. The horizontal red line, with an uncertainty (shaded region) of $\pm$ 4 ns between individual centers, indicates the orbital state lifetime of SiV in unpatterned regions of the chip. Inset: Example decay of the population ratio $A_1/A_2$ of a SiV in a 300 nm pillar.}
\label{fig:sivlifetime}
\end{figure}

To measure the lifetimes of the upper orbital state $\ket{g_2}$ of implanted SiVs in the nanopillars, we optically excited the SiV resonantly on the transition $\ket{g_i}\rightarrow \ket{e_i}$ ($i=1,2$) and monitored the resulting phonon sideband fluorescence as in the resonant excitation measurements from Fig.\ \ref{fig:singlecompare}. We first transfered population from the ground state $\ket{g_1}$ into the upper orbital state via the application of a laser pulse resonant with the transition  $\ket{g_1}\rightarrow \ket{e_2}$, which then decayed into either orbital ground state. By pumping long enough ($\sim$ 80 ns), most population was transferred to state $\ket{g_2}$. Due to this population inversion, the fluorescence collected and measured with an avalanche photodiode decreased during the excitation pulse. To extract the lifetime of the $\ket{g_2}$ state, another 80 ns pulse was applied on the same transition after a wait time $\tau$. Depending on the wait time and the lifetime of the orbital state, some population relaxed back into  ground state $\ket{g_1}$, which manifested itself in an increase in fluorescence compared to the steady state fluorescence at the end of the first pulse. The lifetime $T_1$ could then be extracted by fitting an exponential function
\begin{equation}
\frac{A_{1}}{A_{2}}(\tau)=1+A\ \textrm{exp}\left(-\frac{\tau}{T_1}\right)
\label{eq:sivlt}
\end{equation}
to the ratio of the integrated fluorescence signal during first population inversion pulse $A_1$ and the second  population inversion (readout) pulse  $A_2$. Here $A$ is the maximum observed population inversion fraction. An example spin lifetime measurement of a SiV in a 300 nm pillar is shown in the inset of Fig.~\ref{fig:sivlifetime}.  Note that under resonant excitation the SiV can change its charge state and subsequently turn dark. However, it is possible to reset a SiV to the negative charge state using excitation at 532 nm \cite{meriles2018,Nicolas2019}. Therefore for every 1000 iterations of the life time measurement (1000 $\times$ 2 $\times$ 80 ns, 37 $\mu$W, 737 nm cw power) we applied a charge state reset green laser pulse (45 us, 20 $\mu$W 532 nm cw power). 

Using this method, we measured the orbital state lifetime of implanted SiVs in nanopillars of varying diameter and compared it to the result obtained for SiVs in unpatterned regions of the same chip far from the fabricated pillars. The lifetime of the upper orbital state and its dependence on the pillar diameter is shown in Fig.~\ref{fig:sivlifetime} where the bulk lifetime with an uncertainty of $\pm$ 4 ns (due to the averaging of SiVs in different locations within the bulk) is represented as a horizontal line . 

For SiVs in bulk diamond the lifetime of $\ket{g_{2}}$ is limited by vibrational modes (phonons) \cite{jahnke2015} resonant with the transition to the ground state $\ket{g_{1}}$. However, in small structures the density of these modes is expected to decrease \cite{Lutz2016}.  As the two orbital states $\ket{g_{1,2}}$ that we studied here are spaced by 50 GHz, resonant phonons should be restricted in structures with dimensions smaller than $\approx$ 150 nm. Such a restriction is expected to manifest itself in an increase in the orbital state lifetime. 
Our  measurements show that the largest pillars possess lifetimes comparable to the bulk. In smaller pillars with diameters around 500 nm, some SiVs exhibit slightly longer lifetimes than the bulk average. This may indeed be a weak signature of an onset of the expected lifetime increase in small structures. However, overall our lifetime measurements show a strong scatter with a decreasing trend for smaller pillar sizes.  The strong scatter probably stems from the fact that the measured SiV were not always in the pillar's center since our pillars were fabricated after SiV implantation without first determining the location of individual SiVs. A SiV located closely to the edge of a large pillar where the crystal lattice is imperfect can thus possess a lifetime comparable to a SiV in the center of a much smaller pillar. We believe that reduced lifetimes are caused by imperfections in the crystal lattice surrounding the SiV, likely created during etching. Thus, to confirm our hypothesis that restricted vibrational modes can extend lifetimes, more data, especially from centers that are deterministically placed in high-quality, small pillars would be needed.\\\\

The experiments described in this manuscript show that in order to obtain SiV ensembles with a narrow inhomogeneous broadening it is important to first obtain a high-quality diamond sample with low amounts of residual stress. Even though our diamond samples were nominally identical, two of our three, according to their specifications identical samples displayed a significantly distinct inhomogeneous broadening. Only one sample featured a narrow enough inhomogeneous broadening to resolve the four lines of the SiV. Thus, the available specifications for diamond chips are not sufficient to allow for reproducible SiV fabrication. Re-annealing was not found to be successful in influencing the inhomogeneous linewidth. 
Furthermore, we found that the orbital state lifetime can be affected by nano-fabrication. For some SiVs in pillars with diameters around 500 nm we find slightly increased lifetimes compared to the bulk that may be due to restriction of vibrational modes. However, the measured lifetimes show strong scatter presumably due to random placement of SiVs within the pillars and for the smallest structures we found a trend towards decreased lifetimes; possibly related to imperfections in the crystal lattice or surface damage caused by etching. To confirm an increase in orbital state lifetime due to the suppression of vibrational modes, more experiments with deterministically placed SiVs in high-quality nano pillars with sizes below 500 nm are needed.

\section{Acknowledgements}
Wolfgang Tittel and Paul Barclay thank the Natural Sciences and Engineering Research Council of Canada (NSERC), the Alberta Innovates Technology Futures (AITF) and the Canadian Institute for Advanced Research (CIFAR) for funding. Wolfgang Tittel also thanks the Netherlands Organisation for Scientific Research (NWO).

\end{document}